# Outdoor micro-climate: Air temperature measurements around an office building in Denmark during summer (accepted manuscript for CISBAT 2023)

Hicham Johra[1]*, Mathilde Lenoël[1], Rasmus Lund Jensen[1], Olena Kalyanova Larsen[1]

[1] Department of the Built Environment, Aalborg University, Thomas Manns Vej 23, DK-9220, Aalborg Øst, Denmark

* Corresponding author: hj@build.aau.dk

**Abstract.** The outdoor micro-climate caused by the presence of buildings can significantly differ from that around weather stations located outside of urban areas. However, the latter is often used to design buildings and size building systems. This could lead to significant mistakes and performance gaps. To date, there is a certain lack of experimental studies assessing the micro-climate around buildings, especially in Scandinavian countries. The current paper presents the preliminary results and analysis of a measurement campaign of the temperature gradient in the two-meter air layer around the envelope of a multi-storey office building in Denmark in the summertime. Depending on the orientation of the external building surface (South/North façade or rooftop), the distance from the latter and the weather conditions, the temperature in this two-meter air layer can vary significantly and differ from the air temperature measured at nearby open fields or recorded by the reference weather station. During sunny days, a temperature gradient of up to 3.4 °C and 13.6 °C was measured in the air layer around the South façade and the rooftop, respectively. These results could help to validate urban climate models and bridge the gap between building design and real-condition performance. The curated dataset of this measurement campaign is available in open access.

## 1. Introduction

The outdoor micro-climate induced by the presence of buildings can significantly differ from that around weather stations located outside of urban areas. However, the weather data from these meteorological stations is often used to design buildings' envelope and size heating, cooling and ventilation systems. Such discrepancies can have a severe impact on the actual performance of the systems in direct interaction with the immediate outdoor vicinity of the building: e.g., the capacity of roof-mounted air-source heat pumps and cooling towers during sunny summer days, the effective cooling effect of natural ventilation through windows, or the thermodynamics of double-skin facades.

Currently, many micro-climate studies are based on numerical simulations and focus on the heat island effect in large urban areas subjected to heat waves. Consequently, there is a certain lack of experimental studies assessing the micro-climate around buildings and the difference between rural and urban outdoor conditions, especially in Scandinavian countries. Cold-climate studies mainly focus on winter periods and heating needs. Javanroodi and Nik [1] assess the impact of micro-climate data on wind speed, air temperature, surface temperature, operative temperature and building energy demand for low-density and high-density urban areas. Comparing micro-climate data with Typical Meteorological Year (TMY) data, the operative temperature, surface temperature and winter energy demand can differ by 7%, 27% and 21%, respectively.

Bruelisauer et al. [2] measured the micro-climate temperature around split-type cooling units mounted on the façade of a high-rise building in Singapore. The urban heat island effect is now also studied in colder climates. Toparlar et al. [3] recorded air temperatures in urban areas that are up to 3.3 °C higher than in rural areas for the summer months in Belgium. De Luca et al. [4] concluded that the building cooling demand could be reduced by up to 14% if urban design accounts for micro-climate factors in Nordic countries. Finally, Perini et al. [5] measured the surface and air temperature around green façades during autumn in The Netherlands.

The aim of the present study is to add to the current body of knowledge about outdoor micro-climate around buildings in Scandinavia during the summer period. With the ever-increasing effects of global warming, an accurate estimate of the summer micro-climate becomes crucial for the proper design and operation of buildings, even in cold-climate-dominated countries like Denmark. This article presents the preliminary results and analysis of a measurement campaign of the temperature gradient in the two-meter air layer around the envelope of a multi-storey office building in Denmark in the summertime of 2022. These measurements are compared to far-field temperature measurements, reference meteorological station data and coupled to local weather conditions at the building location.

The curated dataset of this measurement campaign is available in open access [6]. The authors hope it can be useful to the building community that is studying the outdoor micro-climate around buildings in Nordic countries.

## 2. Study case building description

This measurement campaign consists of the monitoring of the outdoor air temperature and weather conditions around the envelope of a multi-storey office building used for teaching and research purposes at the university campus of Aalborg University, Aalborg, Denmark, during the summer of 2022 (see Figure 1). The measurements were carried out from the 10$^{th}$ of June 2022 until the 19$^{th}$ of August 2022.

This study case building is on the western edge of a university campus. It is surrounded by low-rise (two-three floors) office buildings to the North and East. This university campus is situated on the SouthEast edge of the urban area of the city of Aalborg. Aalborg's urban area is of medium population density (178 habitants/km$^2$). It is surrounded by open flatland agricultural fields. It is situated in the north of Denmark, in the north of the Jutland peninsula. Jutland is a low-altitude flatland surrounded by the sea and with strong west-dominated winds. More details on the building study case can be found in a dedicated technical report [7].

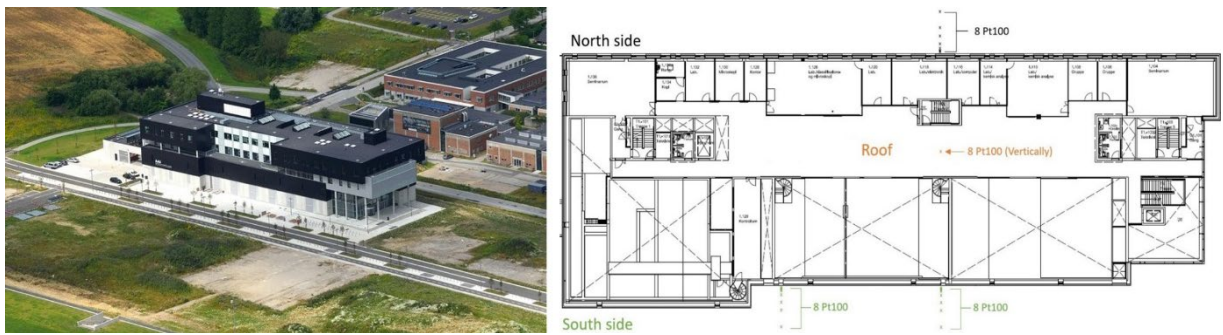

**Figure 1.** Aerial view of the study case building (left). Blueprint of the study case building with the location of the 4 air temperature measurement setups.

The North and South façades are made of light-grey cement slabs on the ground floor (albedo around 0.5) and black and grey metal cover plates on the upper floors. The rooftop is covered with a black asphalt mat (albedo around 0.1).

## 3. Measurement methodology

The temperature in the two-meter air layer around the envelope of the study case building is measured in four locations: North façade, South façade, South façade in front of a ventilation exhaust and on the roof (see Figure 1).

The air temperature is measured at 5 cm, 10 cm, 20 cm, 30 cm, 40 cm, 50 cm, 100 cm and 200 cm from the building envelope. For the North and South façades, the temperature sensors are mounted horizontally at two meters above the ground level (in the centre of the building at the level of light-grey cement slabs constituting the ground floor façades). For the rooftop, the temperature sensors are mounted vertically (see Figure 2).

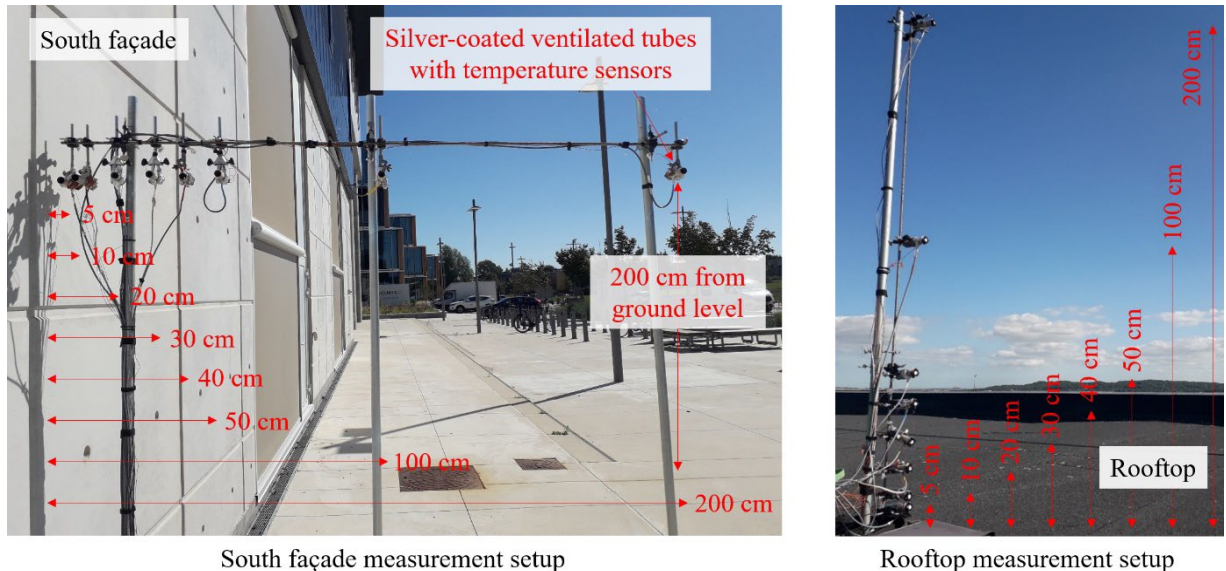

**Figure 2.** Air temperature measurement setups.

The air temperature measurement is performed with calibrated Pt100 resistance temperature detectors connected to NI 9226 cDAQ C Series temperature input modules with a four-wire configuration. To protect the air temperature sensors from solar radiation, the latter are placed inside mechanically-ventilated silver-coated tubes. The uncertainty range for these calibrated temperature sensors is estimated to be ±0.1 K with a 3σ (99.7%) confidence level. Moreover, the outdoor weather condition at the building's location is recorded by a separate meteorological station situated on the rooftop (positioned 5 meters away from the rooftop air temperature measurement setup described above). Weather data is logged continuously and available from a dedicated website: http://www.vejrradar.dk/weatherstation/TMV23/index.php.

The air temperature is also measured at a far-open field (football field) 220 m South of the case building. The air temperature is recorded as the average measurement of 3 calibrated "Tinytag data loggers" placed at two meters height and protected by a ventilated weather station casing. All above-described measurements are logged with a two-second interval and then averaged over five minutes, with the calculation of the standard deviation over that period. Finally, the reference weather data (dry bulb air temperature) is retrieved from the closest reference weather station of the Danish Meteorological Institute (DMI) which is located in Tylstrup, 25 km North of the building case. There is no significant variation in elevation between the four ground stations (at the South façade, the North façade, the far-open field and the DMI station).

More information on the experimental setup used in this study can be found in a dedicated technical report [8]. All raw data from this measurement campaign is open access and can be downloaded [6][8].

## 4. Results and discussion

This section presents and discusses some selected days for which temperature gradients around the building envelope are compared to the far field and the reference meteorological station for different weather conditions.

One can observe in Figure 3 the typical temperature distribution around the building during a sunny afternoon without wind. The air temperature close to the rooftop surface is significantly warmer than the layers above. Although cooler, the air close to the South façade is also very warm. The temperature decreases rapidly when moving away from the heated building surface. At 200 cm from the building envelope, the air temperature is very close to that measured at the far field or the reference DMI weather station. On the other hand, there is no temperature gradient at the North façade, which does not receive solar radiation. The air temperature in the North is slightly lower than that of the far field and the reference DMI weather station. Finally, one can observe a slightly higher standard deviation in the temperature recordings of the rooftop compared to the ground temperature measurements. This temperature variability could be due to a higher air turbulence intensity above the rooftop.

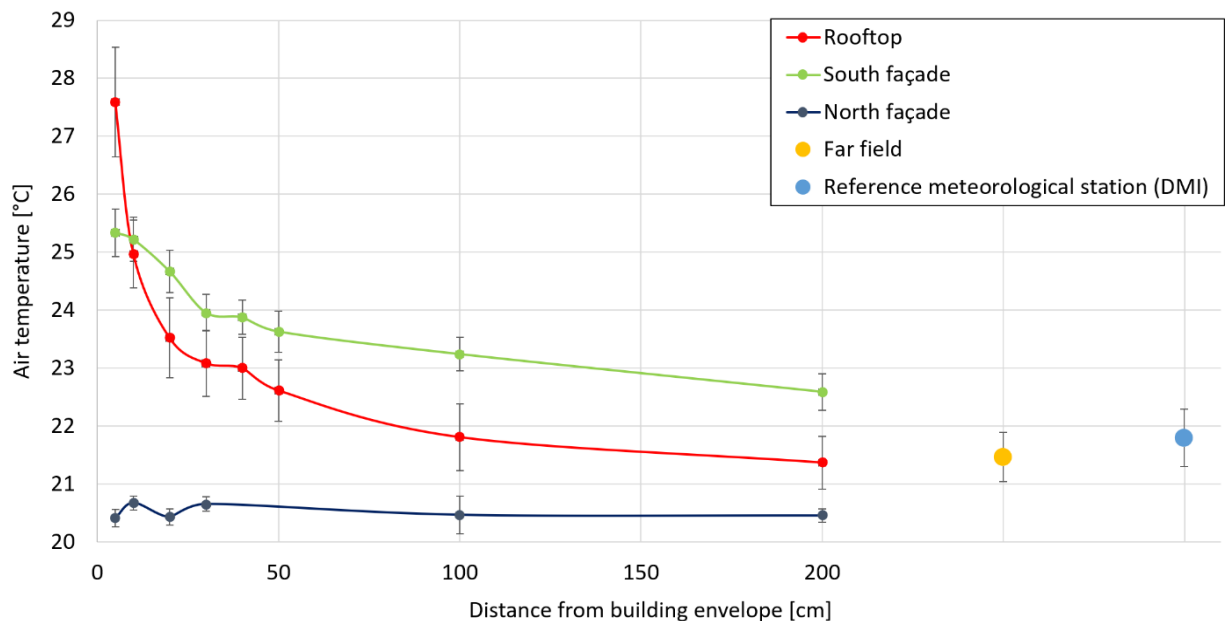

**Figure 3.** Air temperature around the building on a sunny afternoon without wind (average and standard deviation of temperature measurements from 14:00 to 18:00 on the 29[th] of July 2022).

Very similar temperature distribution can be observed for the other sunny afternoons of the summer period, even when it is very windy. In the latter cases, the temperature decreases even faster when moving away from the building envelope. During sunny days, the largest temperature gradient in the air layer (temperature difference between the sensors at 5 cm and 200 cm from the surface) was recorded at 3.4 °C and 13.6 °C for the South façade and the rooftop, respectively.

One can also observe in Figure 4 that the temperature gradient in the air layer on the rooftop increases rapidly when the sun rises and drops rapidly after sunset. This rapid temperature gradient drop might be due to the limited thermal inertia of the roofing layer and the high convection heat exchange induced by the wind. The gradient is insignificant at night. On the rooftop, the temperature gradient even becomes inverted at night (the air temperature close to the roof surface is 2-4 °C colder than the air layers above). This is probably due to the long-wave radiation cooling effect of the sky, especially where there is no cloud cover.

Figure 5 gives an overview of the possible correlations between the air temperature gradient on the rooftop and South façade, and the solar radiation and wind intensity. From the qualitative analysis of this figure, one can observe that a significant linear correlation seems to appear between solar radiation and the temperature gradient: when there is no solar radiation, the temperature gradient is null or very limited. Some outliers can, however, be found at sunset when the solar radiation drops rapidly but the heat accumulated during the day in the rooftop and façade's external layer is released to the surrounding air. On the other hand, it does not seem to be a very distinct correlation between the temperature gradient and the wind speed.

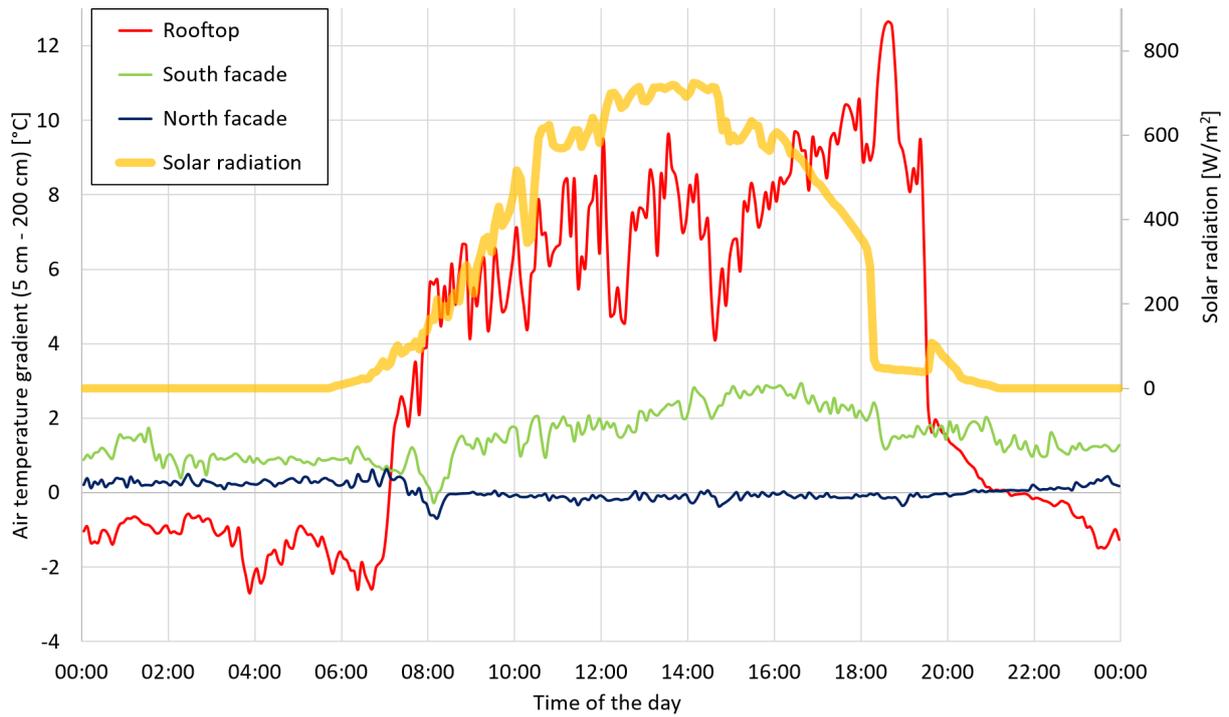

**Figure 4.** Temperature gradient in the air layer around the building (temperature difference between the 5 cm and the 200 cm sensors) on a windy and sunny afternoon: 13[th] of August 2022.

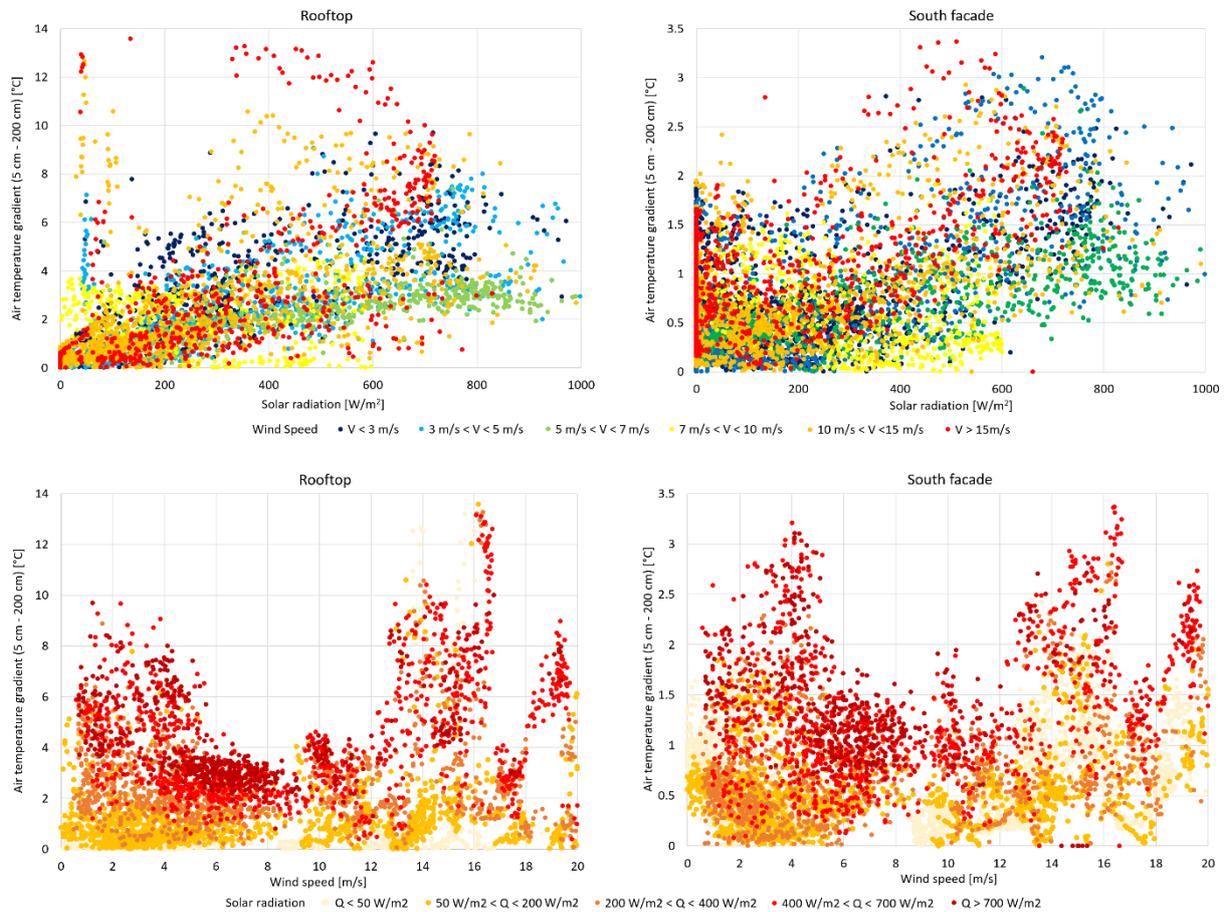

**Figure 5.** Temperature gradient in the air layer on the rooftop and South façade (temperature difference between the 5 cm and the 200 cm sensors) as a function of solar radiation and wind speed.

## 5. Conclusions and future work

This experimental study adds to the body of knowledge on micro-climate around buildings in Scandinavia for summer conditions. The results clearly emphasize the large temperature gradient in the two-meter air layer around the building envelope surfaces exposed to solar radiation during summertime in Denmark. A temperature gradient of up to 3.4 °C and 13.6 °C was measured in the air layer around the South façade and the rooftop, respectively. Moreover, the air close to the building envelope can be close to 10 °C warmer than that of far fields or reference weather stations outside of urban areas.

These significantly warmer air temperatures close to the building envelope might have a significant impact on indoor thermal comfort, energy performance of, e.g., roof-mounted cooling systems, or the cooling effectiveness of natural ventilation through window openings.

Detailed temperature field measurements around buildings can thus foster more accurate design guidelines for building systems with components placed on the external surfaces of the envelope, especially roof-mounted systems. For instance, simple correction factors of reference weather station air temperature located in rural areas or more detailed urban micro-climate models could be beneficial to better estimates the critical summer boundary conditions to size venting, ventilation, cooling and façade systems.

Open empirical datasets, like the one introduced in this article [6], are crucial for the validation and improvement of popular micro-climate models (e.g., ENVI-met) and the development of new ones, especially for countries with very peculiar weather conditions.

The present experimental work continues with further detailed measurements of outdoor micro-climate around different types of buildings in Denmark, coupled with the impact analyses on roof-mounted building systems. All collected results will be described, curated and available in open access.